\documentstyle{elsart}
 
\begin{document}
\begin{frontmatter}

\title{Monte Carlo Hamiltonian -
       From Statistical Physics to Quantum Theory}

\author[ZSU,NCTS,LU]{Xiang-Qian Luo},
\author[ZSU]{C.Q. Huang},
\author[ZSU]{J.Q. Jiang},
\author[LU]{H. Jirari},
\author[LU]{H. Kr\"oger},
\author[DU]{K. J.M. Moriarty}

\address[ZSU]{Department of Physics,
       Zhongshan University, Guangzhou 510275, 
       China\thanksref{Mail}}

\address[NCTS]{National Center for Theoretical Sciences, 
              Hsinchu, Taiwan 30043, China}

\address[LU]{D\'epartement de Physique, Universit\'e Laval, 
Qu\'ebec, Qu\'ebec G1K 7P4, Canada}

\address[DU]{Department of Mathematics, 
Statistics and Computational Science, 
Dalhousie University, Halifax, Nova Scotia B3H 3J5, Canada}

\thanks[Mail]{Mailing Address. E-mail: stslxq@zsulink.zsu.edu.cn}

\begin{abstract}
Monte Carlo techniques have been widely employed in statistical physics
as well as in quantum theory in the Lagrangian formulation.
However, in some areas of application to
quantum theories computational progress has been slow.
Here we present a recently developed approach: the Monte Carlo Hamiltonian
method, designed to overcome the difficulties of the
conventional approach.
\end{abstract}


\end{frontmatter}

\section{Introduction}
The Monte Carlo (MC) method with importance sampling is 
an excellent technique 
to study systems with many degrees of freedom.
It has been widely applied 
to statistical physics and quantum theories.
There are two formulations of quantum theory: 
the Hamiltonian formulation and the Lagrangian formulation. 
A comparison of the conventional approaches is given in Tab. 1.
A natural question is whether one can perform efficiently MC
simulations combined with the Hamiltonian formulation. If so, 
the disadvantages of the Hamiltonian formulation
might be overcome. Our motivation is to construct an effective 
Hamiltonian via MC simulation starting from the original action.

\begin{table}
\caption{Comparison of Hamiltonian and Lagrangian formulations.}
\begin{center}
\begin{tabular}{|c|c|c|}
\hline
{\bf  Formulation} & {\bf Hamiltonian}  &  {\bf Lagrangian} \\
\hline
Approach & Solving Schr\"odinger Eq.    & Computing Path Integral\\
               & $H \vert E_n> = E_n \vert E_n>$
                                          &   $<O> ={\int d[x] O[x]
                                              \exp(- S[x]/ \hbar)  \over
                                     \int [dx] \exp(- S[x]/ \hbar)}$ \\
\hline
Algorithm & variational, series expansion,  & Monte Carlo simulation\\
          &  coupled cluster expansion ...  & as in {\bf Stat. Phys.}\\
\hline
Advantage  & can obtain not only energies  & use computer to generate \\
           & of ground $\&$ excited states, & the most important configs.\\ 
           & but also wavefunctions   & obeying Boltzmann's law      \\
\hline
Disadvantage & methods above are too      & difficult to obtain: wavefunction,\\
             & tedious for realistic    & excited states, S matrix,\\
             & applications                  & finite density QCD\\
\hline
            
\end{tabular}
\end{center}
\end{table}


\section{Algorithm}
Let us review briefly the basic ideas \cite{mch}. 
According to Feynman's path integral approach to quantum mechanics, 
the (imaginary time) transition amplitude between 
an initial state at position  
$x_i$, and time $t_i$,  
and final state at $x_f$, $t_f$ is related 
to the Hamiltonian $H$ by
\begin{eqnarray}
&& <x_{f},t_{f} | x_{i},t_{i}> 
= <x_{f} | e^{-H(t_{f}-t_{i})/\hbar} | x_{i}>
\nonumber \\
&& = \sum_{\nu=1}^{\infty} <x_{f} | E_{\nu} > 
e^{-E_{\nu} T/\hbar} < E_{\nu} | x_{i}>,
\end{eqnarray}
where $T=t_f-t_i$.
The starting point of our method, as described in more detail in \cite{mch}
is to construct an effective Hamiltonian $H_{eff}$ 
(finite $N \times N$ matrix) by 
\begin{eqnarray}
&& <x_{f} | e^{-H(t_{f}-t_{i})/\hbar} | x_{i}> \approx 
<x_{f} | e^{-H_{eff} T/\hbar} | x_{i}>
\nonumber \\
&& =\sum_{\nu=1}^{N} < x_{f} | E^{eff}_{\nu}> e^{-E^{eff}_{\nu} T/\hbar} 
< E^{eff}_{\nu} | x_{i} > .
\end{eqnarray}
$H_{eff}$ can be found by MC simulation 
using the following procedure: 

\noindent
(a) Discretize the continuous time.

\noindent
(b) Generate configurations $[x]$ obeying the Boltzmann distribution
\begin{eqnarray}
P(x) = { \exp(- S[x]/ \hbar)  \over
       \int [dx] \exp(- S[x]/ \hbar)}.
\end{eqnarray}
\noindent
(c) Calculate the transition matrix elements 
\begin{eqnarray}
M_{fi} = 
<x_{f} | e^{-H_{eff} T/\hbar} | x_{i}>
\end{eqnarray}
between $N$ discrete $x_i$ points and $N$ $x_f$ points. 
Note that the matrix $M$ is symmetric.

\noindent
(d) Diagonalize $M$ by a unitary transformation
\begin{eqnarray}
M &=& U^{\dagger}DU,
\end{eqnarray}
where 
$D =diag (e^{-E^{eff}_{1}T/\hbar},..., e^{-E^{eff}_{N}T/\hbar})$. 
Steps (a) and (b) are the same as the standard MC method.
Step (c) is the essential ingredient of our method, from which
we can construct $H_{eff}$, and obtain
the eigenvalues $E^{eff}_{\nu}$ and wavefunction 
$\vert E^{eff}_{\nu} >$ 
through step (d). Once the spectrum
and wave functions are available, all physical information 
can be retrieved.
Since the theory described by $H$ is now approximated by a theory
described by a finite matrix $H_{eff}$, the physics
of $H$ and $H_{eff}$ might be quite different at high energy. 
Therefore we expect that we can only
reproduce the low energy physics of the system.
This is good enough for our purpose.

\section{Testing the Method}
We have tested a number of quantum mechanical models in both 1+1 dimensions 
\cite{mch,1d,lat99_2}
and 2+1 dimensions \cite{lat99_1,2d}.
Here we will show some new results for a model 
in 1+1 dimensions with the following potential
\begin{equation}
V(x)=x^2/2+x^4/4.
\end{equation}
The Euclidean action is given by 
$S = \int_{0}^{T} dt [m \dot{x}^{2}/2 + V(x)]$.  
This model is not exactly solvable.
We have compared the results from our method with those 
from a standard algorithm (Runge-Kutta algorithm
plus node theorem).
The first three eigenvalues are given in Tab. 2. 
\begin{table}
\caption{Spectrum of the model with 
$V(x)=x^2/2+x^4/4$. $E_n^{n.t.}$ corresponds to Runge-Kutta plus node theorem and $\Delta E_n^{M.C.}$ is the 
statistical error.}
\vskip 3mm
\begin{center}
\begin{tabular}{|c|c|c|c|}
\hline
n & $E_n^{n.t.}$& $E_n^{M.C.}$  & $\Delta E_n^{M.C.}$\\
\hline
0  &  0.6209 & 0.6197 & 0.0192\\
1  &  2.0260 & 2.0427 & 0.0566\\
2  &  3.6985 & 3.6060 & 0.0455\\
\hline
\end{tabular}
\end{center}
\vskip 3 mm
\end{table}
The parameters are $m=1$, $T=1$, $\hbar=1$, $\Delta x=1$, $N=21$.
The results for the first two wave functions are shown in Figs. 1 and 2.
Besides the eigenvalues $E^{eff}_{\nu}$ and wave functions 
$\vert E^{eff}_{\nu} >$, 
we have computed thermodynamical quantities 
such as the partition function $Z$,
average energy $\overline{E}$ and specific heat $C$. 
Since we have approximated $H$ by $H_{eff}$, 
we can express those thermodynamical observables  
via the eigenvalues of the effective Hamiltonian
\begin{eqnarray}
Z(\beta) &=& \sum_{\nu=1}^{N}e^{-\beta E_{\nu}^{eff}}
\nonumber \\
\overline{E}(\beta) &=& \sum_{\nu=1}^{N}
{E_{\nu}^{eff}e^{-\beta E_{\nu}^{eff}} \over Z(\beta)},
\nonumber \\
C(\beta) &=& k_B{\beta}^2\left(\sum_{\nu=1}^{N}
{(E_{\nu}^{eff})^2e^{-\beta E_{\nu}^{eff}} \over
Z(\beta)}-\overline{E}^2(\beta) \right),
\end{eqnarray}
where $\beta=T/\hbar$.  
The results for the average energy and specific heat 
are shown in Figs. 3 and 4,
where the solid lines corresponds to the standard algorithm.
Within statistical errors, our results are consistent with those from
the standard algorithm.

\begin{figure}
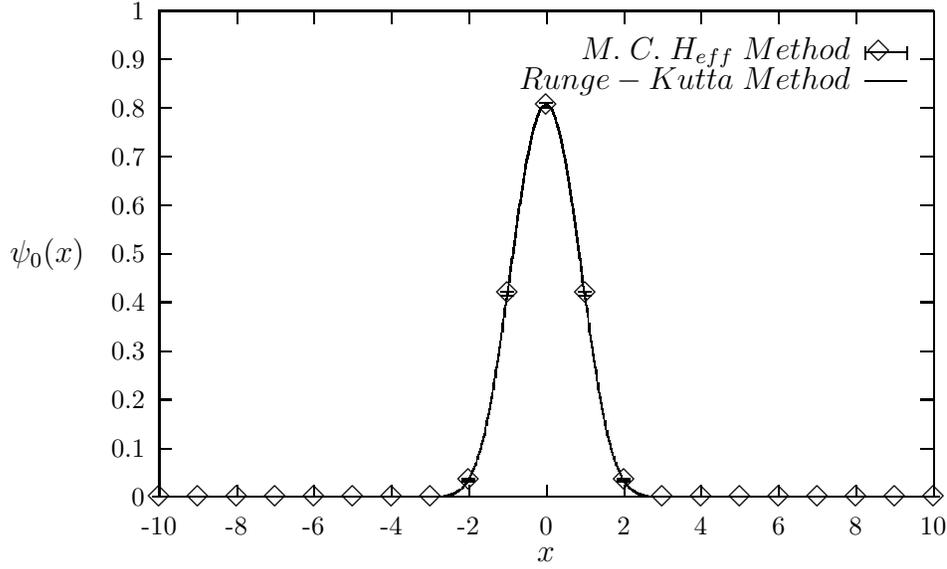

\begin{center}
\setlength{\unitlength}{0.240900pt}
\ifx\plotpoint\undefined\newsavebox{\plotpoint}\fi

\end{center}

\caption{Ground state wave function of the model with 
$V(x)=x^2/2+x^4/4$.}
\end{figure}



\begin{figure}
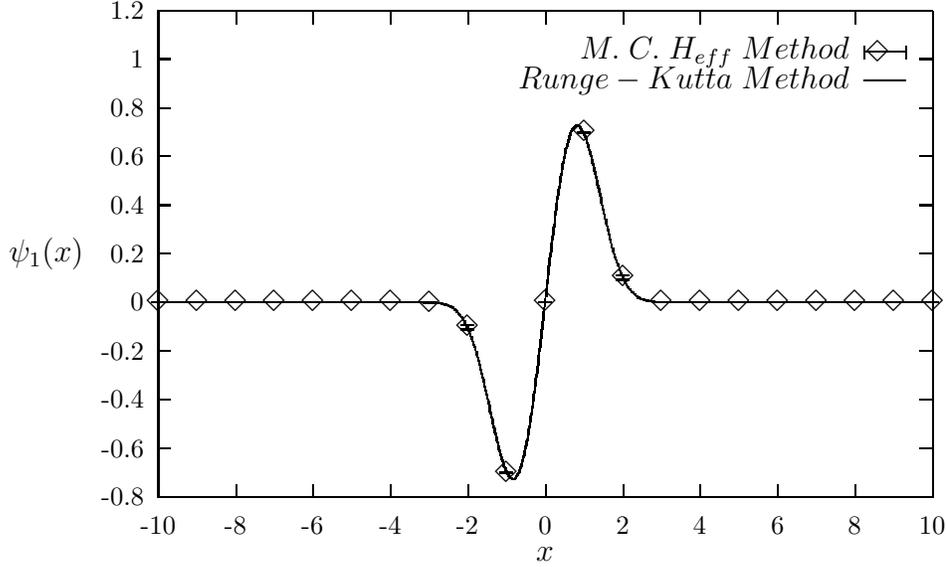


\begin{center}
\setlength{\unitlength}{0.240900pt}
\ifx\plotpoint\undefined\newsavebox{\plotpoint}\fi
%
\end{center}
\caption{First excited state of the model with $V(x)=x^2/2+x^4/4$.}
\end{figure}


\begin{figure}
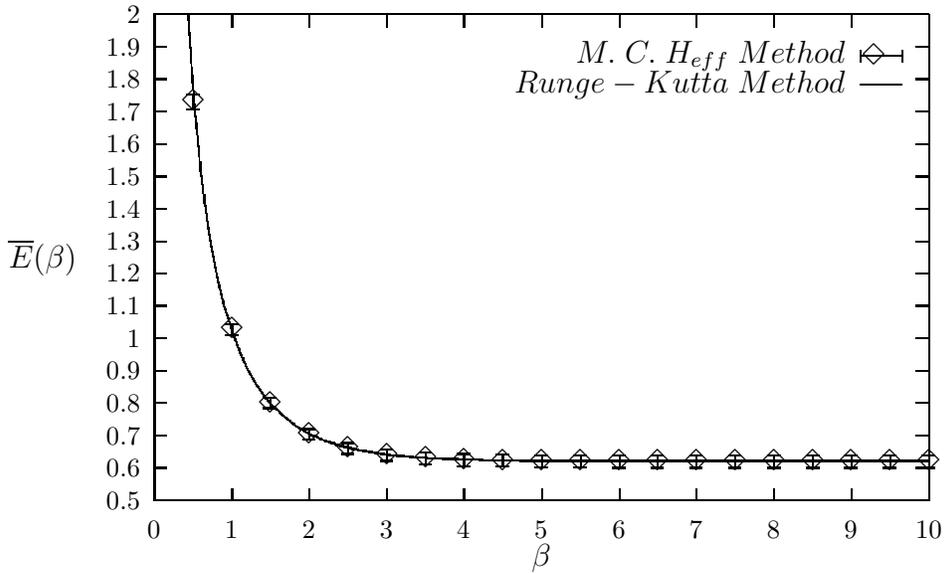

\setlength{\unitlength}{0.240900pt}
\ifx\plotpoint\undefined\newsavebox{\plotpoint}\fi
%
\caption{Average Energy of the model with $V(x)=x^2/2+x^4/4$.}
\end{figure}

\begin{figure}
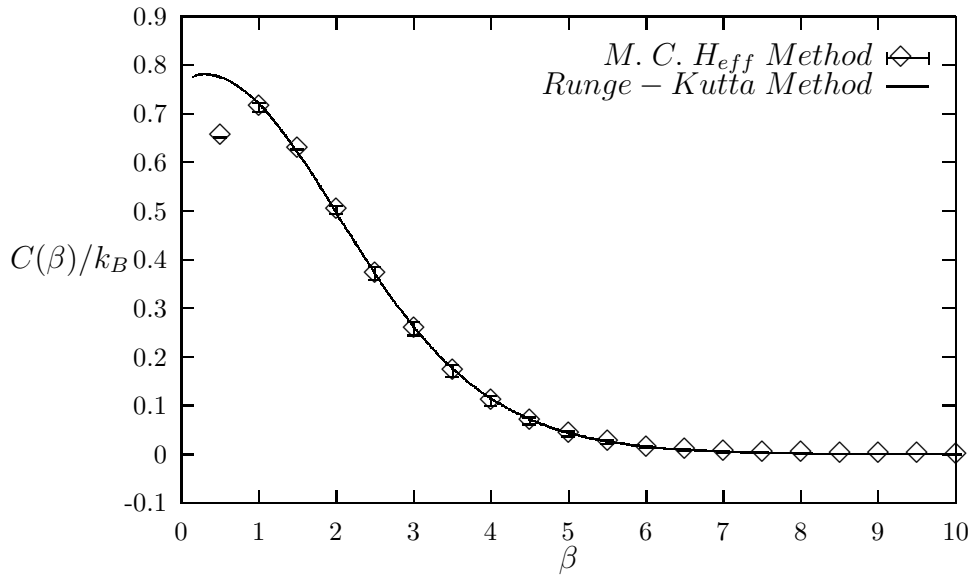

\setlength{\unitlength}{0.240900pt}
\ifx\plotpoint\undefined\newsavebox{\plotpoint}\fi
%
\caption{Specific heat of the model with $V(x)=x^2/2+x^4/4$.}
\end{figure}

\section{Summary and Outlook}
We have presented the basic ideas and some new results
from our recently proposed Monte Carlo effective Hamiltonian method.
We found that the method works well also for the $V \sim x^{4}$ potential,
allowing to compute excited states and thermodynamical observables.
In the previous simulations, 
the initial and final positions $x_i$ and $x_f$ were chosen 
to be uniformly distributed.
As can be seen, for a given dimension of the Hilbert space $N$, 
the results break down if $\beta$ is too small.
Increasing $N$ requires larger CPU time. 
For quantum mechanics in 3+1 dimensions or quantum field theories, 
a stochastic basis will be necessary to select
the most important contributions to the transition matrix $M_{fi}$.
Such a work is in progress \cite{lat99_2}.

\ack

X.Q.L. is supported by the
National Science Fund for Distinguished Young Scholars,
National Natural Science Foundation, 
the Chinese Ministry of Education, 
Hong Kong Foundation of
the ZSU Advanced Research Center,
and NCTS, Taiwan.
H.K. and K.J.M.M. have been supported by NSERC Canada.


\end{document}